\documentclass[aps,prb,twocolumn,superscriptaddress,showpacs]{revtex4}
\usepackage{color}
\usepackage{graphicx}
\usepackage{cancel}
\usepackage{amsfonts,amsmath}
\def  \bsig    {\mbox{\boldmath$\sigma$}}

\begin{document}
\bibliographystyle{apsrmp}

\title{Thermoelectric effect enhanced by the resonant states in graphene}
\author{M. Inglot}
\affiliation{Department of Physics, Rzesz\'ow University of Technology, Al.~Powsta\'nc\'ow Warszawy 6, 35-959 Rzesz\'ow, Poland}
\author{A. Dyrda\l}
\affiliation{Faculty of Physics, Adam Mickiewicz University, ul. Umultowska 85, 61-614 Pozna\'n, Poland}
\author{V. K. Dugaev}
\affiliation{Department of Physics, Rzesz\'ow University of Technology, Al.~Powsta\'nc\'ow Warszawy 6, 35-959 Rzesz\'ow, Poland}
\affiliation{Departamento de F\'isica and CFIF, Instituto Superior T\'ecnico,
Universidade de Lisboa, Av.~Rovisco Pais, 1049-001 Lisbon, Portugal}
\author{J. Barna\'s}
\affiliation{Faculty of Physics, Adam Mickiewicz University, ul. Umultowska 85, 61-614 Pozna\'n, Poland}
\affiliation{Institute of Molecular Physics, Polish Academy of Sciences,
ul. M. Smoluchowskiego 17, 60-179 Pozna\'n, Poland}

\date{\today }

\begin{abstract}
Thermoelectric effects in graphene are considered theoretically with particular attention paid to the role of impurities. Using the $T$-matrix
method we calculate the impurity resonant states and the momentum relaxation time due to scattering on impurities. The Boltzmann kinetic equation is used to determine the thermoelectric coefficients. It is shown that the resonant impurity states near the Fermi level give rise to a resonant enhancement of
the Seebeck coefficient and of the figure of merit $ZT$. The Wiedemann-Franz ratio deviates from that known for ordinary metals, where this ratio is constant and equal to the Lorentz number. This deviation appears for small chemical potentials and in the vicinity of the resonant states. In the limit of a constant relaxation time, this ratio has been calculated analytically for $\mu =0$.
\end{abstract}

\maketitle
\section{Introduction}

Graphene, the first strictly two-dimensional crystal, has been extensively studied, both experimentally
and theoretically. This enormous interest in graphene results mainly from peculiar transport properties, which follow from its very specific electronic structure. Thermal and thermoelectric properties of graphene have become a topical issue since the first measurement of thermal conductivity in graphene by Balandin {\it et al}.~\cite{balandin2008} It is known that graphene is one of the best heat conductors, with a very high thermal conductivity that is a consequence of the strong $sp^{2}$ bonding, light atomic mass, and  low dimensionality.~\cite{Xu2014} A giant thermoelectric effect in graphene was also predicted theoretically,~\cite{Dragoman} with the Seebeck coefficient equal to 30 mV/K.

Impurities can substantially influence graphene's energy spectrum, so the thermoelectric transport strongly depends not only on the thermal activation but also on the impurity scattering. The influence of impurity scattering on the thermoelectric properties of graphene was considered theoretically using a self-consistent t-matrix approach by Lofwander {\it et al}.\cite{Lofwander} From these considerations follows that the  measured thermopower  can be used to get some information on the role of impurity scattering in graphene. The thermopower near the Dirac point was considered by Wang {\it et al},\cite{Wang} who obtained a relatively good agreement between experimental data and theoretical results based on the Boltzmann transport theory. These authors also showed that the Mott's relation fails in the vicinity of the Dirac points in the case of high-mobility graphene, being however satisfied for a wide range of gate voltages  in the regime of  low carrier mobility. The thermoelectric transport in graphene was also considered for Dirac fermions in the presence of magnetic field and disorder ~\cite{Zuev,Zhu,Yan}.

Sharapov {\it et al}~\cite{Sharapov} have found that the thermopower can be remarkably enhanced by opening an energy gap in the quasiparticle spectrum of graphene. The presence of energy gap is accompanied by the emergence of quasiparticle scattering channel, with the relaxation time strongly dependent on energy.
The thermoelectric effects ware also investigated in multilayer graphene systems. The thermopower of biased and unbiased multilayer graphene was studied in the Slonczewski-Weis-McClure model, where the effect of impurity scattering was treated in the self-consistent Born approximation\cite{hao}.  The classic and spin Seebeck effects in single as well as in multilayer graphene on the SiC substrate were also investigated within ab-initio methods,~\cite{Wierzbowska1,Wierzbowska2} and by nonequilibrium molecular-dynamics simulations.~\cite{fthenakis}

Transport properties of graphene nanoribbons (GNRs) can differ from the corresponding ones of a two-dimensional graphene plane,
mainly due to edge sates and energy gaps which develop in the electronic spectrum. These transport properties also depend on the
edge shape.
Thermoelectric properties of GNRs have also been investigated and it has been
shown that thermopower in GNRs can be remarkably larger
than that in the planar graphene.\cite{wei} Moreover, the corresponding figure
of merit, $ZT$, can be enhanced by introducing randomly
distributed hydrogen vacancies into completely hydrogenated
GNRs.\cite{liang} Structural defects, especially in the form of antidots, were also shown to be
a promising way of enhancing thermoelectric efficiency in
GNRs.\cite{gunst,karamitaheri,yan,chang,wierzbicki}

In this paper we consider thermoelectric properties of a two-dimensional graphene with impurities which lead to
resonant states near the Fermi level. We show that the resonant states result in resonant enhancement of the
Seebeck coefficient. This enhancement can be observed when the Fermi level is tuned by an external gate voltage. In section  2 we describe the model and theoretical method used to calculate the Seebeck coefficient.
The relaxation time is calculated in section 3. Numerical results on the thermoelectric transport
properties are presented and discussed in section 4. Wiedemann-Franz law is briefly discussed in section 5, wile final conclusions are in section 6.

\section{Model}

In a clean (defect-free) graphene, the electronic states in the vicinity of the Dirac points can be described by the
following low-energy Hamiltonian \cite{kane}
\begin{equation}
\hat H_{0} = -i \hbar v \, (\sigma_{x} \nabla _{x} + \sigma_{y} \nabla _{y})\, ,
\end{equation}
where $v$ is the electron velocity in graphene,
and $\sigma_x$, $\sigma _y$ are the Pauli matrices defined
in the two-sublattice space of graphene.
This Hamiltonian describes two electron energy bands with linear dispersion,
$\varepsilon ^{(1,2)}({\bf k})=\pm \hbar vk$. The electron velocity in each of the bands is
$\mathcal{\bf v}^{(1,2)}({\bf k})=\pm v{\bf k}/\  k$.

Assume that a temperature gradient $\nabla T$
and external electric field $E$ are oriented along the axis $x$.
Both, $\nabla T$ and $E$ drive the system out of equilibrium.
To calculate the distribution function $f^{(n)}({\bf r,k})$ of
electrons in the $n$-th energy band in the nonequilibrium situation,
we apply the Boltzmann kinetic equation.
For a weak deviation $\delta f^{(n)}$ of the distribution function
from the equilibrium distribution $f_0$,
the Boltzmann equation
for $\delta f^{(n)}$ can be written in the relaxation time approximation as
\begin{eqnarray}
\label{2}
v^{(n)}_x\, \left(-\frac{\partial f_0}{\partial \varepsilon }\right)\,
\left(\nabla \mu +\frac{\varepsilon -\mu }{T}\; \nabla T
-eE \right) =-\frac{\delta f^{(n)}}{\tau ^{(n)}}\, ,
\end{eqnarray}
where $f^{(n)}({\bf r,k})=f_0+\delta f^{(n)}$,
$\tau^{(n)}$ is the relaxation time in the $n$-th band,
while $\mu $ is the chemical potential, which may be spatially inhomogeneous along the axis $x$ and
thus $\nabla \mu=\partial\mu /\partial x$ is a driving force as well.

Using the solution of Eq.~(2), one can find the electric current density $j$ and the energy flux density
$J_E$ along the axis $x$, induced by the driving forces $E$, $\nabla T$, and $\nabla \mu$,\cite{kireev}
\begin{eqnarray}
\label{3}
j=e\sum _{n{\bf k}}v^{(n)}_x\delta f^{(n)}=
e^2K_{11}E  \hskip 1cm
\nonumber \\
-eK_{11}T\, \nabla\frac{\mu }{T}-eK_{21}\frac{\nabla T}{T}\, ,
\end{eqnarray}
\begin{equation}
J_E=\sum _{n{\bf k}}v^{(n)}_x\varepsilon ^{(n)}\, \delta f^{(n)}
=eK_{21}-K_{21}T\, \nabla\frac{\mu }{T}-K_{31}\frac{\nabla T}{T}\, ,
\end{equation}
where $K_{rs}$ are the kinetic coefficients for graphene
\begin{equation}
\label{5}
K_{rs}=-\frac1{4\pi \hbar ^2}\int _{-\infty }^\infty |\varepsilon |\, \varepsilon ^{r-1}
\tau ^s(\varepsilon )\, \frac{\partial f_0}{\partial \varepsilon }\, d\varepsilon .
\end{equation}
The integral in Eq.(5) runs over both energy bands, so
$\tau (\varepsilon )$ is equal to $\tau ^{(1)}(\varepsilon )$ for $\varepsilon >0$,
and $\tau ^{(2)}(\varepsilon )$ for $\varepsilon <0$, respectively.
Note, that the heat current density $J_Q$ is defined as $J_Q=J_E-\mu j$, so that $J_Q=J_E$ when $j=0$.

The thermoelectric Seebeck coefficient $\alpha $ and the heat conductivity $\kappa $ are defined from
the relations $E^\alpha =\alpha\nabla T$ and $J_Q=-\kappa \nabla T$ when $j=0$, where
$E^\alpha $ is the electric field due to the temperature gradient. In general, there is also a field
 related to the chemical potential inhomogeneity, $E^\mu =\nabla \mu /e$,
so that the total internal electric field for $j=0$ is $E=E^\mu +E^\alpha $.
This leads to the standard expressions for the electrical conductivity $\sigma$, thermoelectric Seebeck coefficient $\alpha$, and
heat conductance $\kappa$~\cite{kireev}
\begin{eqnarray}
\label{6}
&&\sigma =e^2K_{11},
\\
&&\alpha =\frac{K_{21}-\mu K_{11}}{e\, TK_{11}}
\\
&&\kappa =\frac{K_{31}K_{11}-K_{21}^2}{TK_{11}},
\end{eqnarray}
which are valid for graphene when the kinetic coefficients $K_{rs}$ are
calculated from Eq.~(5).

To calculate the thermoelectric parameters, Eqs(6-8), it is necessary to know the energy dependence of
the relaxation time due to electron scattering from impurities and defects. This problem is considered in the section below.

\section{Momentum relaxation time}

The total Hamiltonian of graphene with impurities is $\hat{H}=\hat{H}_0+\sum _i \hat{V}({\bf r-R}_i)$,
where $\hat{V}({\bf r -R_i})$ is a scattering potential of a single impurity located at ${\bf R}_i$,
and the sum goes over all randomly distributed impurities.
We consider the situation, when the short-range-potential impurities are distributed randomly with
equal probabilities in the sublattices A and B of the graphene. Correspondingly, the single-impurity
perturbation  is either $\hat{V}^A({\bf r})$ or $\hat{V}^B({\bf r})$, where
\begin{equation}
\label{9}
\hat V^{A,B}({\bf r})=\hat{V}_0^{A,B}\, \delta({\bf r}-{\bf R}_{A,B}),
\end{equation}
where $\hat{V}_0^{A,B}=V_0 (\sigma_{0}\pm \sigma_z)/2$,
$V_0$ is the impurity potential strength, $\sigma_0$ is the unit matrix in the sublattice space and
${\bf{R}}_{A}$ (${\bf{R}}_{B}$) is a position vector of the impurity if it is in the sublattice A (B).

The influence of a single impurity on the energy spectrum and on the momentum relaxation time
can be described in terms of the T-matrix method~\cite{mahan}.
The T-matrix equation in the general case takes the form
\begin{equation}
\label{10}
\hat T_{\bf k k'}(\varepsilon) = \hat V_{\bf k k'}
+\sum_{\bf k''} \hat V_{\bf k k''} \hat G_{\bf k''}(\varepsilon)\, \hat T_{\bf k'' k'}.
\end{equation}
where
\begin{eqnarray}
\label{11}
\hat{G}_{\bf k}(\varepsilon )=\frac{\varepsilon +\bsig \cdot {\bf k}}{(\varepsilon +i\delta )^2-(\hbar vk)^2}
\end{eqnarray}
is the retarded Green's function of electrons in graphene described by the Hamiltonian $\hat{H}_0$.

In the case of short-range impurities, we can find from Eq.~(10) the T-matrix in the following
form
\begin{equation}
\label{12}
\hat T^{A,B}(\varepsilon)=\frac{\hat{V}^{A,B}_{0}}{1-V_{0}F(\varepsilon)}\, ,
\end{equation}
where $F(\varepsilon )$ is defined as
\begin{eqnarray}
\label{13}
F(\varepsilon )=\sum _{\bf k} \frac{\varepsilon }{(\varepsilon +i\delta )^2-(\hbar vk)^2}\, .
\end{eqnarray}

Using Eq.~(12), and  averaging over the impurity
positions (assuming the same probability to find impurity in the sublattice
A or B), we can find the self-energy of electrons due to the scattering from
impurities
\begin{eqnarray}
\label{14}
\hat{\Sigma }(\varepsilon )=\frac{N_iV_0\, \sigma _0}{2\, [1-V_0F(\varepsilon )]}\, ,
\end{eqnarray}
where $N_i$ is the impurity concentration.

To find the relaxation time of electrons in the energy bands 1 and 2, we have to diagonalize
the operator $\hat{H}_{eff}\equiv \hat{H}_0+\hat{\Sigma }(\varepsilon )$.
Since the self energy (14) is proportional to the unit matrix $\sigma _0$,
the operator $\hat{H}_{eff}$ is diagonalized by the same transformation as
$\hat{H}_0$. In other words, the relaxation time of electrons in graphene
can be found directly from Eq.~(14) by using relation
\begin{eqnarray}
\label{15}
\frac{\hbar }{\tau (\varepsilon )}
={\rm Im}\; \frac{N_iV_0}{1-V_0 F(\varepsilon )}\, .
\end{eqnarray}
It can be also presented as
\begin{equation}
\label{16}
\frac{\hbar }{\tau(\varepsilon)}
=\frac{N_iV_0^2\, |{\rm Im}\, F(\varepsilon )|}
{[1-V_0{\rm Re}\, {F}(\varepsilon )]^2 + V_0^2[{\rm Im}\, {F}(\varepsilon )]^2}\, ,
\end{equation}
where the real and imaginary parts of $F(\varepsilon )$ can be calculated from Eq.~(13)
\begin{align}
\label{17}
&{\rm Re}\, F(\varepsilon )
\simeq -\frac{\varepsilon }{2\pi (\hbar v)^2}\;
\ln \frac{\hbar vk_m}{|\varepsilon |}\, ,
\\
&
{\rm Im}\, F(\varepsilon )\simeq
-\,
\frac{\varepsilon }{4(\hbar v)^2}\, ,
\end{align}
and $k_m$ is a maximum value of the wave vector (cutoff) in graphene, $k_m=(|{\bf K}|+|{\bf M}|)/2$,
with ${\bf K}$ and ${\bf M}$ denoting the corner and edge of graphene BZ measured from the
$\Gamma$ point.

\section{Numerical results: thermopower and figure of merit $ZT$}

Using Eqs.(3)-(5) we calculate first the electric current for
$E=0$ and homogeneous chemical potential, $\nabla \mu =0$.
The current is then solely induced by the temperature gradient,
$j=\sigma \alpha \nabla T$.
Figure 1(a) presents the thermoelectric current  as a function of
the chemical potential $\mu $, which experimentally can be tuned by an external gate voltage.
In the calculations we used the following parameters:
$\hbar v=1.05\times 10^{-28}$~Jm, $k_m=1.59\times 10^{10}$~m$^{-1}$,
$N_i=2\times 10^{14}$~m$^{-2}$, temperature $T=300$~K, and $\nabla T=8000$~K/m.
As follows from Fig.1, the magnitude of current as well as its variation with the chemical
potential $\mu$ strongly depend on the impurity potential $V_0$.

\begin{figure}[t]
\centering\includegraphics[width=0.99\columnwidth]{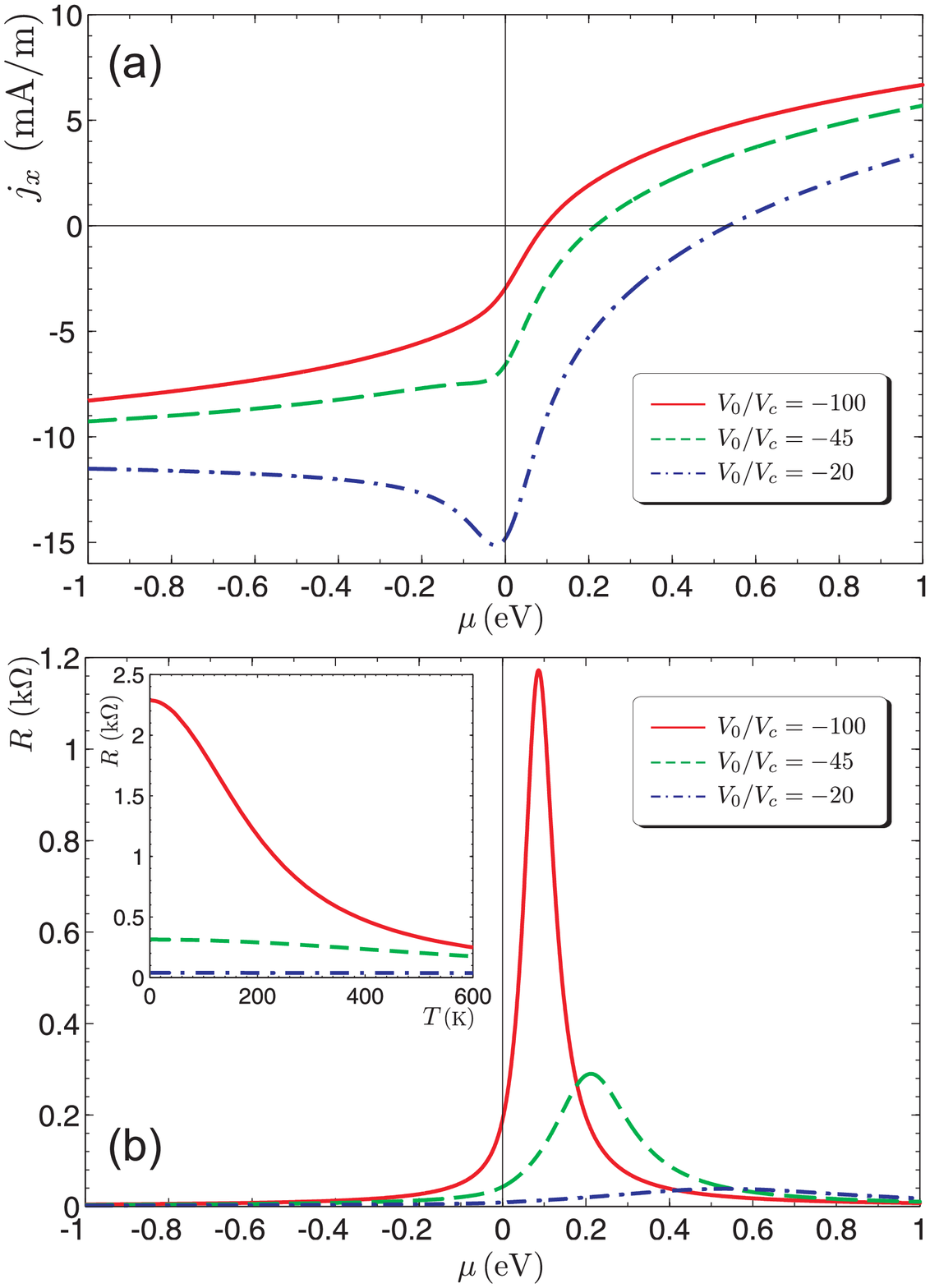}
\caption{\label{fig1}(Color online) (a) Current $j_x$ induced by a
temperature gradient $\nabla T$ as a function of the chemical potential $\mu$
for different values of the impurity potential $V_0$. (b)
Resistivity of graphene vs $\mu $ for different $V_0$. The inset in (b) shows the resistivity {\it vs} temperature.
The parameters assumed as described in the main text.}
\label{jt}
\end{figure}

The  dependence of the thermoelectric current on $\mu $ is closely related to
the presence of resonance states. This can be also seen from the resistivity behavior,
which has a pronounced maximum when the chemical potential is located in the
vicinity of a resonance impurity state, see Fig.1(b). The location of the resonance states is
determined by the magnitude of the impurity potential $V_0$.\cite{inglot11}
To understand behavior of the thermocurrent one should note that particles and holes
flow from higher temperature to the lower one (from right to left for the assumed temperature gradient).
Figure 1 shows that the thermocurrent vanishes at the chemical potential,
where the resistance achieves a maximum value.
At this point the current due to
electrons is compensated by the current due to holes, so the total current vanishes.
In turn, for higher chemical potentials the particle current dominates so the total current is positive, while
for lower chemical potentials (left of the resistance maximum) the hole current dominates and current is negative.

\begin{figure}[t]
\centering\includegraphics[width=0.99\columnwidth]{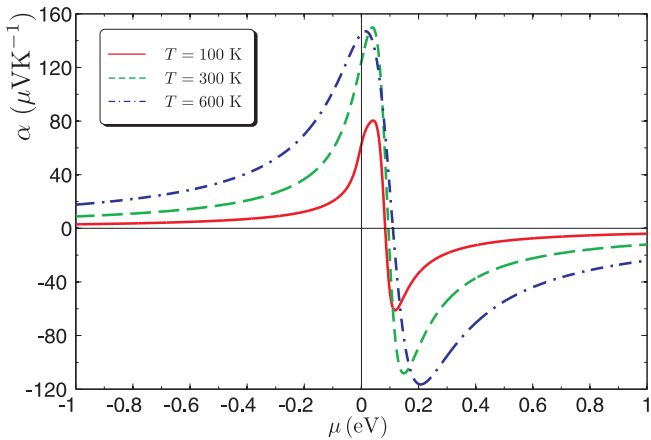}
\caption{\label{fig2}(Color online) Seebeck coefficient $\alpha$ of graphene
as a function of chemical potential $\mu$ for indicated values of  temperatures $T$.}
\label{jt}
\end{figure}

The above described behavior of the thermocurrent and electrical resistivity
is also reflected in
the dependence of the Seebeck coefficient $\alpha $ on the chemical potential $\mu$,
as shown in Fig.~2. To calculate $\alpha $,  we used Eq.~(7) and take the parameters
$V_0/V_c=-100$ and $N_i=2\times 10^{14}$~m$^{-2}$, where
$V_c=ta_0^2$, $t=3$~eV is the hopping integral and $a_0=1.42\times 10^{-10}$~m is
the distance between carbon atoms in graphene. Obviously, the thermopower (Seebeck coefficient) is
equal to zero at the chemical potentials where the thermocurrent vanishes. For larger chemical potentials, the thermopower is negative since the current is dominated by particles. In turn, for lower
chemical potentials the thermopower is positive as the current is then dominated by holes. Interestingly,
the maxima in the absolute magnitude of the thermopower appear at the points, where
the change in resistance (and thus also in
transmission through the graphene) with the chemical potential reaches a maximum.

\begin{figure}[t]
\centering\includegraphics[width=0.99\columnwidth]{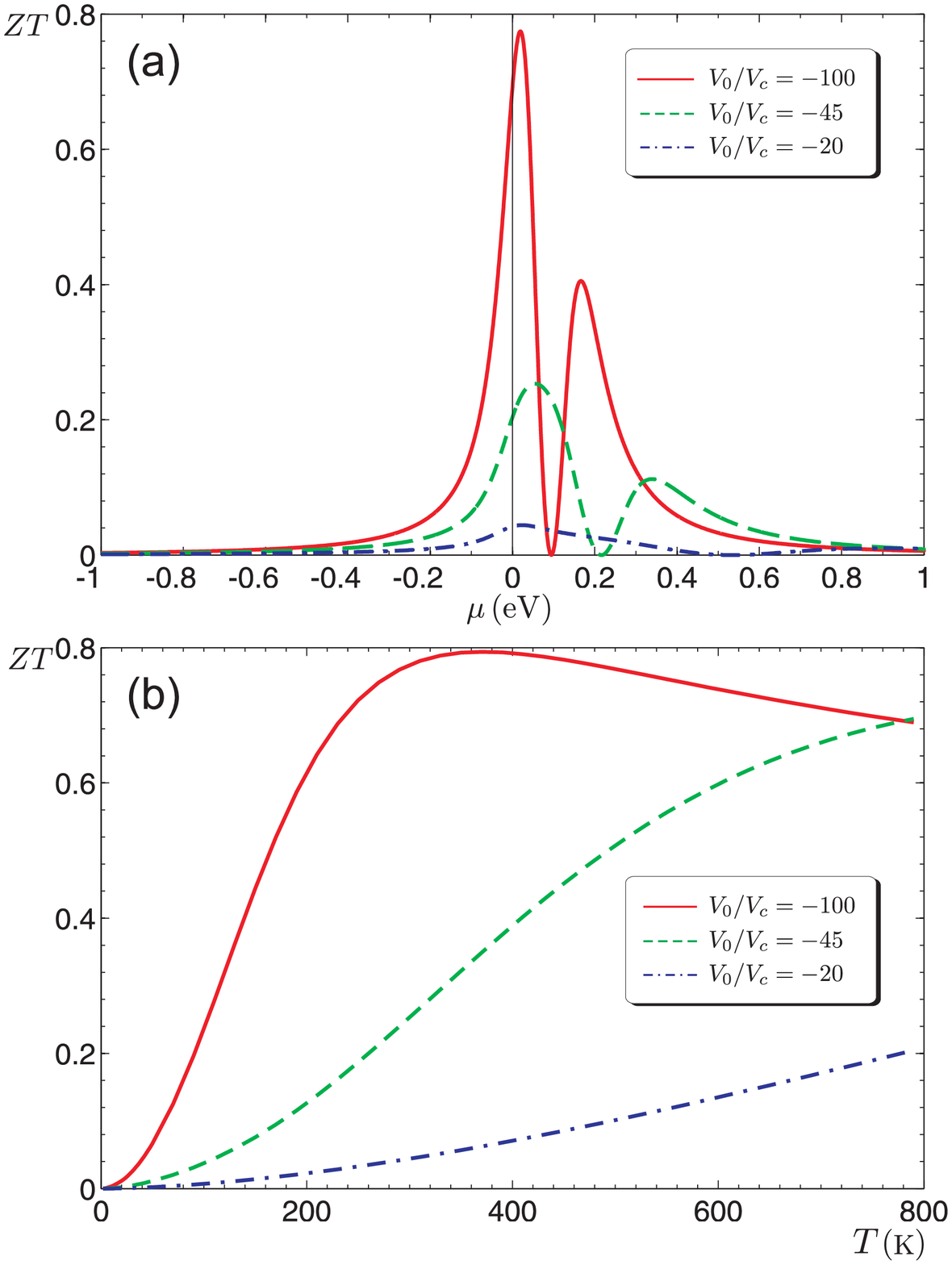}
\caption{\label{fig3}(Color online) Figure of merit $ZT$ for graphene as a
function of the chemical potential $\mu$ (a) and as function of temperature $T$ (b),
for indicated values of the impurity potential
$V_0$.}
\label{jt}
\end{figure}	

Figure 3 shows the corresponding figure of merit $ZT$, defined as
\begin{align}
\label{19}
ZT = \frac{T\alpha^2\sigma}{\kappa },
\end{align}
as a function of the chemical potential, Fig.~3(a), and as a function of
temperature, Fig.~3(b). The temperature dependence of $ZT$ in Fig. 3(b) is shown for
the chemical potentials $\mu $, where the corresponding $ZT$ achieves a maximum in Fig.~3(a).
Note, the figure of merit vanishes at the chemical potentials where the thermopower vanishes (maxima of the resistance).
However, there is a significant enhancement of the
figure of merit due to resonant scattering on impurities, and the enhancement appears at the chemical potential where
the electrical resistance varies rapidly with $\mu$.
However, one should bear in mind, that this figure of merit
has been calculated taking into account the electronic term in the heat conductance only.
A phonon contribution to the heat conductance inevitably suppresses
$ZT$ to remarkably smaller values.
	
\begin{figure}[t]
\centering\includegraphics[width=0.99\columnwidth]{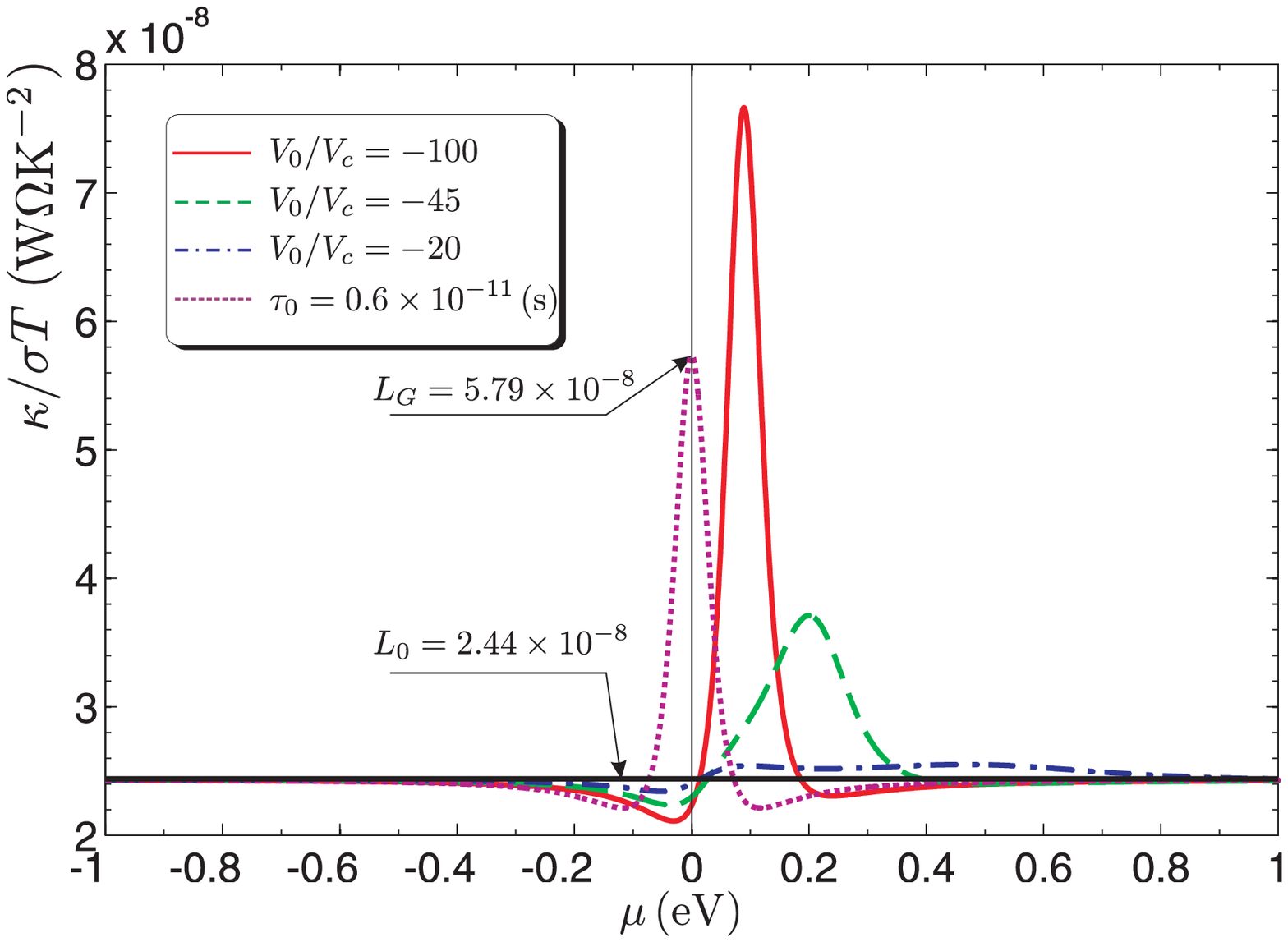}
\caption{\label{fig4}(Color online) The Wiedemann-Franz ratio
as a function of chemical potential $\mu$ for $T=300$ K. The black
line corresponds to the Lorentz number in metals, $L_0=2.44 \times 10^{-8}\
{\rm W\Omega K^{-2}}$, whereas $L_G=5.79 \times 10^{-8}$~W${\rm \Omega }$K$^{-2} $ corresponds to
the Wiedemann-Franz ratio for graphene at $\mu =0$ and constant relaxation time.}
\end{figure}

\section{The Wiedemann-Franz law for graphene}

Let us consider now the Wiedemann-Franz law. In ordinary metals this law states that the ratio $\kappa/ \sigma T$ is constant,
\begin{align}
\label{20}
\frac{\kappa}{T\sigma} = L\, ,
\end{align}
where $L=L_0= \pi ^2 k_B^2/3e^2 =2.44\times 10^{-8}\ {\rm W \Omega K}^{-2}$ is a constant, known as the Lorentz number.
The situation in graphene is different. Strong energy dependence of the relaxation time and the presence of resonance states,
lead to a significant dependence of the ratio $\kappa/ \sigma T$ on the chemical potential $\mu$, as shown in Fig.~4 for $T=300$~K and indicated values of the scattering potential $V_0$.
Experimentally, the chemical potential can be tuned by an external gate voltage.
As one can see, this ratio is not constant, but strongly depends on the chemical potential $\mu $, especially
in the vicinity of the resonance states. Moreover,  this ratio also depends on the impurity potential $V_0$.
However, far from the resonance states, when the Fermi level is well inside the upper (positive $\mu$) or lower (negative $\mu$) band,
the ratio $\kappa/ \sigma T$ tends to the number typical in metals, i.e. to $L_0$. It is rather clear, that the deviation of the ratio $\kappa/ \sigma T$ from $L_0$ follows from the resonance states and specific electronic structure of graphene (energy dependent density of states).
Similar deviations are also known in other systems,
for instance in transport through nanoscopic quantum objects like quantum dots, molecules, etc.

It is interesting to consider the situation which may be directly compared to ordinary metals, i.e. when the
relaxation time due to impurity scattering
is constant $\tau =\tau _0$ in the vicinity
of the point $\mu =0$.  The corresponding ratio $L$ is shown in figure 4.
When the chemical level is well inside the  valence or conduction bands, $|\mu |>>0$, the ratio
$\kappa/ \sigma T$  tends to the Lorentz number $L_0$, as in the case of resonance states. However, this ratio
has a maximum at $\mu =0$, i.e. at the point where the density of states in graphene disappears.

One can calculate the Wiedemann-Franz ratio of graphene at $\mu =0$ analytically
assuming that the relaxation time is constant $\tau =\tau _0$ in the vicinity
of the point $\mu =0$. In other words, we assume that one can neglect the dependence of $\tau $
on energy, $\delta \tau (\varepsilon )/\tau \ll 1$ for $|\varepsilon |\leq k_BT$,
which is certainly fulfilled at rather low temperatures for any dependence
$\tau (\varepsilon )$.
In such a case, the kinetic coefficient $K_{21} = 0$, and
the other coefficients can be calculated analytically
\begin{eqnarray}
\label{21}
&&K_{11}=\frac{\tau _0k_B T\ln 2}{2\pi \hbar^2} \, ,
\\
&&K_{31}=
\frac{9\zeta (3)\tau _0 k_B^3T^3}{4\pi \hbar^2} \, .
\end{eqnarray}
Using Eq.(21) and Eq.(22), one can calculate the Lorentz number for graphene at $\mu =0$,
\begin{equation}
\label{23}
L_G\equiv \left. \frac{\kappa}{\sigma T}\right| _{\mu =0}
=\frac{9k_B^2\, \zeta(3)}{2e^2\ln 2}\, ,
\end{equation}
where $\zeta (x)$ is the Riemann's $\zeta $-function.
This value of $L_{G}$ is equal to $5.795070903\times 10^{-8}\, {\rm W \Omega K^{-2}}$.
Taking into account that $L = \pi ^2 k_B^2/3e^2$ we can also find
\begin{equation}
\label{24}
\frac{L_G}{L}
=\frac{27\, \zeta(3)}{2\pi^2 \ln 2}\simeq 2.37.
\end{equation}
This result was also obtained earlier by Saito et al.\cite{saito2007} for the case of purely
ballistic electric and thermal conductance of graphene, when $\mu =0$.

\section{Conclusions}

We have analyzed theoretically the thermoelectric properties of graphene with
impurities equally distributed in both A and B sublattices.
These impurities lead to resonance impurity states near the Fermi level, and therefore to resonance electron scattering.
In agrement with the Mott's law, the magnitude of the Seebeck coefficient $\alpha$
and shape of its dependence on the chemical potential is strongly affected by the dependence
of the conductivity on $\mu $, which in turn is determined mainly by the
electron scattering  from impurities in the vicinity of resonance states.
The resonant states also lead to significant enhancement of the figure of merit $ZT$.

We have also shown that the ratio $\kappa/ \sigma T$ deviates from the Lorentz number $L_0$, and this deviation
appears for chemical potentials in the vicinity of the resonant states. Moreover, assuming a constant relaxation time we
have found analytical solution for the ratio $\kappa/ \sigma T$ at $\mu=0$, which agrees with that derived earlier.

In our calculations we introduced the impurity density $N_i$, the impurity scattering potential $V_0$,
and the chemical potential $\mu $ as independent parameters. This assumption
can be justified, if there are other (not necessarily resonant) impurities and defects
in graphene. Additionally, the chemical potential can be tuned also by an external gate voltage, which
gives an additional experimental tool to study the thermoelectric properties
of graphene and their dependence on these parameters.

\begin{acknowledgments}
This work was supported by the National Science Center in Poland as research
projects Nos. DEC-2011/01/N/ST3/00394 (MI), DEC-2012/06/M/ST3/00042 (VKD), and DEC-2012/04/A/ST3/00372 (AD,JB).
\end{acknowledgments}


\end{document}